%% file: ms.tex
\documentclass[preprint]{aastex631} 

\usepackage{url}
\usepackage{hyperref}
\usepackage{xspace}
\usepackage{color}

\newcommand{\Fermi}{\emph{Fermi}\xspace}
\newcommand{\Swift}{\emph{Swift}\xspace}
\newcommand{\Suzaku}{\emph{Suzaku}\xspace}
\newcommand{\INT}{\emph{INTEGRAL}\xspace}
\newcommand{\AGILE}{\emph{AGILE}\xspace}
\newcommand{\RHESSI}{\emph{RHESSI}\xspace}

\newcommand{\Insight}{\emph{Insight}\xspace}
\newcommand{\MO}{\emph{Mars-Odyssey}\xspace}
\newcommand{\MES}{\emph{MESSENGER}\xspace}
\newcommand{\Wind}{\emph{Wind}\xspace}

\begin{document}

\title{The Second Catalog of Interplanetary Network Localizations of Konus Short Duration Gamma-Ray Bursts}

\author{D. Svinkin}
\affiliation{Ioffe Institute, Politekhnicheskaya 26, St. Petersburg, 194021, Russia}
\author{K. Hurley}
\affiliation{Space Sciences Laboratory, University of California, 7 Gauss Way, Berkeley, CA 94720-7450, USA}
\author{A. Ridnaia}
\affiliation{Ioffe Institute, Politekhnicheskaya 26, St. Petersburg, 194021, Russia}
\author{A. Lysenko}
\affiliation{Ioffe Institute, Politekhnicheskaya 26, St. Petersburg, 194021, Russia}
\author{D. Frederiks}
\affiliation{Ioffe Institute, Politekhnicheskaya 26, St. Petersburg, 194021, Russia}
\author{S. Golenetskii}
\affiliation{Ioffe Institute, Politekhnicheskaya 26, St. Petersburg, 194021, Russia}
\author{A. Tsvetkova}
\affiliation{Ioffe Institute, Politekhnicheskaya 26, St. Petersburg, 194021, Russia}
\author{M. Ulanov}
\affiliation{Ioffe Institute, Politekhnicheskaya 26, St. Petersburg, 194021, Russia}
\author{A. Kokomov}
\affiliation{Ioffe Institute, Politekhnicheskaya 26, St. Petersburg, 194021, Russia}
\author{T.~L.~Cline}
\affiliation{13708 Sherwood Forest Drive, Silver Spring, MD 20904, USA}

\author{I.~Mitrofanov}
\affiliation{Space Research Institute, 84/32 Profsoyuznaya, Moscow, 117997, Russia}
\author{D.~Golovin}
\affiliation{Space Research Institute, 84/32 Profsoyuznaya, Moscow, 117997, Russia}
\author{A.~Kozyrev}
\affiliation{Space Research Institute, 84/32 Profsoyuznaya, Moscow, 117997, Russia}
\author{M.~Litvak}
\affiliation{Space Research Institute, 84/32 Profsoyuznaya, Moscow, 117997, Russia}
\author{A.~Sanin}
\affiliation{Space Research Institute, 84/32 Profsoyuznaya, Moscow, 117997, Russia}

\author{A.~Goldstein} 
\affiliation{Science and Technology Institute, Universities Space Research Association, Huntsville, AL 35805, USA}
\author{M.S.~Briggs}
\affiliation{Space Science Department, University of Alabama in Huntsville, 320 Sparkman Drive, Huntsville, AL 35899, USA}
\author{C.~Wilson-Hodge} 
\affiliation{NASA Marshall Space Flight Center, Huntsville, AL 35812, USA}
\author{E.~Burns} 
\affiliation{Department of Physics \& Astronomy, Louisiana State University, Baton Rouge, LA 70803, USA}

\author{A.~von~Kienlin} 
\affiliation{Max-Planck-Institut f\"{u}r extraterrestrische Physik, Giessenbachstrasse 1, D-85748 Garching, Germany}
\author{X.-L.~Zhang} 
\affiliation{Max-Planck-Institut f\"{u}r extraterrestrische Physik, Giessenbachstrasse 1, D-85748 Garching, Germany}
\author{A.~Rau} 
\affiliation{Max-Planck-Institut f\"{u}r extraterrestrische Physik, Giessenbachstrasse 1, D-85748 Garching, Germany}
\author{V. Savchenko}
\affiliation{Department of Astronomy, University of Geneva, Ch. d'Ecogia 16, 1290, Versoix, Switzerland}
\author{E.~Bozzo} 
\affiliation{Department of Astronomy, University of Geneva, Ch. d'Ecogia 16, 1290, Versoix, Switzerland}
\author{C.~Ferrigno}
\affiliation{Department of Astronomy, University of Geneva, Ch. d'Ecogia 16, 1290, Versoix, Switzerland}

\author{S. Barthelmy}
\affiliation{NASA Goddard Space Flight Center, 8800 Greenbelt Road, Greenbelt, MD 20771, USA}
\author{J.~Cummings} 
\affiliation{Center for Astrophysical Sciences, Johns Hopkins University, Baltimore, MD, USA}
\author{H.~Krimm} 
\affiliation{National Science Foundation, Alexandria, VA 22314, USA}
\author{D.~M.~Palmer}
\affiliation{Los Alamos National Laboratory, B244, Los Alamos, NM 87545, USA}
\author{A.~Tohuvavohu}
\affiliation{Department of Astronomy and Astrophysics, University of Toronto, Toronto, ON, Canada}

\author{K. Yamaoka} 
\affiliation{Institute for Space-Earth Environmental Research(ISEE), Nagoya University, Furo-cho, Chikusa-ku, Nagoya 464-8601, Japan} 
\author{M. Ohno}
\affiliation{Institute of Physics, E{\"o}tv{\"o}s University, P{\'a}zm{\'a}ny P{\'e}ter s{\'e}t{\'a}ny 1/A, Budapest, 1117, Hungary} 
\affiliation{School of Science, Hiroshima University, 1-3-1 Kagamiyama, Higashi-Hiroshima, Hiroshima 739-8526, Japan} 
\affiliation{Konkoly Observatory of the Hungarian Academy of Sciences, Konkoly-Thege ut 15-17, Budapest, 1121, Hungary} 
\author{Y. Fukazawa}
\affiliation{Department of Physics, Hiroshima University, 
1-3-1 Kagamiyama, Higashi-Hiroshima, Hiroshima 739-8526, Japan}
\author{Y. Hanabata}
\affiliation{Department of Physics, Hiroshima University, 
1-3-1 Kagamiyama, Higashi-Hiroshima, Hiroshima 739-8526, Japan}
\author{T. Takahashi}
\affiliation{Department of Physics, The University of Tokyo, 7-3-1 Hongo, Bunkyo, Tokyo 113-0033, Japan} 
\affiliation{Kavli Institute for the Physics and Mathematics of the Universe (WPI), 
The University of Tokyo Institutes for Advanced Study, The University of Tokyo, Kashiwa, Chiba 277-8583, Japan} 
\author{M. Tashiro}
\affiliation{Department of Physics, Saitama University, 
255 Shimo-Okubo, Sakura-ku, Saitama-shi, Saitama 338-8570, Japan}
\author{Y. Terada}
\affiliation{Department of Physics, Saitama University, 
255 Shimo-Okubo, Sakura-ku, Saitama-shi, Saitama 338-8570, Japan}
\author{T. Murakami}
\affiliation{Department of Physics, Kanazawa University,
Kadoma-cho, Kanazawa, Ishikawa 920-1192, Japan}
\author{K. Makishima}
\affiliation{Department of Physics, University of Tokyo, 7-3-1
Hongo, Bunkyo-ku, Tokyo 113-0033, Japan}

\author{W.~Boynton}
\affiliation{Lunar and Planetary Laboratory, University of Arizona, Tucson, AZ, USA}
\author{C.W.~Fellows} 
\affiliation{Lunar and Planetary Laboratory, University of Arizona, Tucson, AZ, USA}
\author{K.P.~Harshman}
\affiliation{Lunar and Planetary Laboratory, University of Arizona, Tucson, AZ, USA}
\author{H. Enos}
\affiliation{Lunar and Planetary Laboratory, University of Arizona, Tucson, AZ, USA}
\author{R. Starr}
\affiliation{Catholic University of America, Washington, DC 20064, USA}

\author{J. Goldsten}
\affiliation{Applied Physics Laboratory, Johns Hopkins University, Laurel, MD 20723, USA}
\author{R. Gold}
\affiliation{Applied Physics Laboratory, Johns Hopkins University, Laurel, MD 20723, USA}

\author{A. Ursi}
\affiliation{INAF/IAPS, via del Fosso del Cavaliere 100, I-00133 Roma (RM), Italy}
\author{M. Tavani}
\affiliation{INAF/IAPS, via del Fosso del Cavaliere 100, I-00133 Roma (RM), Italy}
\author{A. Bulgarelli}
\affiliation{INAF/OAS, via Gobetti 101, I-40129 Bologna (BO), Italy}
\author{C. Casentini} 
\affiliation{INAF/IAPS, via del Fosso del Cavaliere 100, I-00133 Roma (RM), Italy}
\author{E. Del~Monte} 
\affiliation{INAF/IAPS, via del Fosso del Cavaliere 100, I-00133 Roma (RM), Italy}
\author{Y. Evangelista}
\affiliation{INAF/IAPS, via del Fosso del Cavaliere 100, I-00133 Roma (RM), Italy}
\author{M. Galli} 
\affiliation{INAF/OAS, via Gobetti 101, I-40129 Bologna (BO), Italy}
\affiliation{ENEA Bologna, via don Fiammelli 2, I-40128 Bologna (BO), Italy}
\author{F. Longo}
\affiliation{Dipartimento di Fisica, Universit\`a di Trieste and INFN, via Valerio 2, I-34127 Trieste (TR), Italy}
\author{M. Marisaldi} 
\affiliation{INAF/OAS, via Gobetti 101, I-40129 Bologna (BO), Italy}
\affiliation{Birkeland Centre for Space Science, Department of Physics and Technology, University of Bergen, Norway}
\author{N. Parmiggiani} 
\affiliation{INAF/OAS, via Gobetti 101, I-40129 Bologna (BO), Italy}
\author{C. Pittori} 
\affiliation{SSDC/ASI, via del Politecnico snc, I-00133 Roma (RM), Italy}
\affiliation{INAF/OAR, via Frascati 33, I-00078 Monte Porzio Catone (RM), Italy}
\author{M. Romani}
\affiliation{INAF - Osservatorio Astronomico di Brera, Via Brera, 28 - I-20121 Milano (MI), Italy}
\author{F. Verrecchia}
\affiliation{SSDC/ASI, via del Politecnico snc, I-00133 Roma (RM), Italy}
\affiliation{INAF/OAR, via Frascati 33, I-00078 Monte Porzio Catone (RM), Italy}

\author{D.~M.~Smith}
\affiliation{Physics Department and Santa Cruz Institute for Particle Physics, 
University of California, Santa Cruz, Santa Cruz, CA 95064, USA}
\author{W.~Hajdas}
\affiliation{Paul Scherrer Institute, 5232 Villigen PSI, Switzerland}

\author{S.~Xiao}
\author{C.~Cai}
\author{Q.~B.~Yi} 
\author{Y.~Q.~Zhang} 
\author{S.~L.~Xiong}
\author{X.~B.~Li}
\author{Y.~Huang}
\author{C.~K.~Li}
\author{S.~N.~Zhang}
\author{L.~M.~Song}
\author{C.~Z.~Liu}
\author{X.~Q.~Li}
\author{W.~X.~Peng}
\affiliation{Key Laboratory of Particle Astrophysics, Institute of High Energy Physics, 
Chinese Academy of Sciences, 19B Yuquan Road, Beijing 100049, China}

\author{I.~Martinez-Castellanos}
\affiliation{Department of Physics, University of Maryland, College Park, MD, USA}

\begin{abstract}
We present the catalog of Interplanetary Network (IPN) localizations for 199 short-duration gamma-ray bursts (sGRBs)
detected by the Konus-\textit{Wind} (KW) experiment between 2011 January~1 and 2021 August~31,
which extends the initial sample of IPN localized KW sGRBs~\citep{Palshin13} to 495 events.  
We present the most comprehensive IPN localization data on these events, 
including probability sky maps in HEALPix format. 
\end{abstract}

\section{INTRODUCTION} \label{sec:intro} 
Between 1994 November and 2021 August, the Konus-\Wind gamma-ray spectrometer (KW; \citealt{Aptekar95}) 
on board the \emph{Global Geospace Science Wind} spacecraft detected 
3394 gamma-ray bursts (GRBs) in the triggered mode, 495 of which were classified as short-duration
gamma-ray bursts (sGRBs) or short bursts with extended emission (EE); see \cite{Svinkin16,Svinkin19} for the KW short/long GRB classification criteria.

Here we present the localization data obtained by arrival time analysis, or ``triangulation'' 
between the spacecraft in the third interplanetary network (IPN) for 199 sGRBs that occurred 
during the period from 2011 January~1 to 2021 August~31.
The IPN localizations for 296 KW sGRBs detected in 1994-2010 were presented earlier (\citealt{Palshin13}, hereafter~P13).
Due to KW's continuous coverage (duty cycle $\gtrsim 95$\%) of the full sky by two omni-directional detectors over a wide energy range ($\sim 20$--15~MeV), 
the KW sample is the most complete set of sGRBs with fluences above $\sim 10^{-6}$~erg~cm$^2$~s$^{-1}$ available to date.

The short GRB sample is not homogeneous; it includes both Type~I (merger-origin)
and Type~II (collapsar-origin) GRBs; see \citealt{Zhang_2009ApJ_703_1696} 
for more information on the Type I/II classification scheme.
Taking into account the burst durations and hardness ratios~\citep{Svinkin16,Svinkin19} we estimate that
about 20\% of the bursts in our sample can be in fact Type~II, 
or at least their classification as Type~I is questionable.
The sample also includes three possible short GRBs with EE.

Recently, rapid IPN localizations have facilitated significant discoveries in the GRB field, 
e.g. the localization of the short GRB~170817A~--- the counterpart of the gravitational wave event GW~170817 
from a binary neutron star merger~\citep{Abbott_2017ApJ_848L_12},
the detection of the extragalactic magnetar giant flare (MGF) in NGC~253~\citep{Svinkin_2021Natur}, 
and the discovery and confirmation of the shortest gamma-ray burst from a collapsar~\citep{Ahumada_2021NatAs}.
This catalog provides essential information for gravitational wave and neutrino searches from short GRBs,
and for searches for extragalactic MGFs.

The novel feature of this catalog is the presentation of the final GRB localizations 
as probability sky maps using the Hierarchical Equal Area isoLatitude Pixelization (HEALPix\footnote{\url{https://healpix.sourceforge.io}}) 
discretization~\citep{Gorsky_2005ApJ_622_759, Zonca_2019JOSS_4_1298}.
Since HEALPix has been recently accepted as a standard data format for multi-messenger astronomy, such localizations will 
aid the joint analysis of localizations involving gravitational wave observatories, 
as well as ground- and space-based facilities across the electromagnetic spectrum.

This catalog is organized as follows.
In Section~\ref{sec:obs} we describe the composition of the IPN in 2011--2021
and briefly discuss the instrumentation and GRB observations. 
In Section~\ref{sec:localizations} we provide triangulation annuli, 
other localization constraints used, and the methodology of deriving IPN error regions.
Section~\ref{sec:loc_res} presents the final IPN error regions and discusses the statistics of the localizations;
in Section~\ref{sec:conclusions} we give our conclusions.
A description of the data used and localization file format can be found in 
Appendices~\ref{appendix:data} and~\ref{appendix:loc}, respectively.
All coordinates are aberration-corrected equinox J2000.

\section{OBSERVATIONS}\label{sec:obs}

The composition of the missions and experiments comprising the IPN changes as old missions are terminated and new missions are introduced. 
During the period covered in the present catalog (2011-2021), the IPN contained between 7 and 9 missions: 
Konus-\Wind, in orbit around the Lagrangian point $L_1$;
\MO (the Gamma-Ray Spectrometer, GRS, that includes the High Energy Neutron Detector, HEND, with GRB detection capabilities; \citealt{Hurley06}), 
in orbit around Mars; 
the Mercury Surface, Space Environment, Geochemistry, and Ranging mission (\MES; the Gamma-Ray and Neutron Spectrometer, GRNS; \citealt{Gold01}), 
in an eccentric orbit around Mercury; 
the International Gamma-Ray Laboratory (\INT; the anti-coincidence shield of the spectrometer SPI, SPI-ACS), in an eccentric Earth orbit \citep{Rau05}; 
the Ramaty High Energy Solar Spectroscopic Imager (\RHESSI; the array of germanium detectors, GeD; \citealt{Smith02});
the Neil Gehrels \Swift Observatory (the Burst Alert Telescope, BAT; \citealt{Gehrels04});
the \Fermi Gamma-ray Space Telescope (the Gamma-Ray Burst Monitor, GBM; \citealt{Meegan09});
the \Suzaku mission (the Wide-band All-sky Monitor, WAM; \citealt{Takahashi07,Yamaoka09});
the Astro-rivelatore Gamma a Immagini LEggero mission (\AGILE; the Mini-Calorimeter, MCAL; \citealt{Tavani09}); 
the CALorimetric Electron Telescope aboard the \emph{International Space Station} (CALET; Gamma-ray Burst Monitor;  \citealt{Yamaoka_2013ICRC}); 
the Hard X-ray Modulation Telescope (\Insight-HXMT; the High Energy X-ray telescope, HE; \citealt{Zhang_2020SCPMA_6349502}); and
the Gravitational Wave high-energy Electromagnetic Counterpart All-sky Monitor 
(GECAM; the Gamma-Ray Detectors, GRD; \citealt{Chen_2021GECAM}; 
GECAM consists of two microsatellites: GECAM-A and GECAM-B, but currently only GECAM-B is in operation); all in low Earth orbit.
Table~\ref{tab:instruments} lists the operation period, the distance from the Earth, the time resolution, the energy range of 
the detector used for triangulation, and the number of KW short GRBs observed by each mission/instrument.

For each KW sGRB we searched for detections in the data of the IPN spacecraft
taking into account the possible range of propagation time delays. 
For each instrument we searched for a corresponding trigger or waiting mode detection (if available from the instrument team).
For CALET we used triggered events reported via the Gamma-ray Coordinates Network\footnote{\url{https://gcn.gsfc.nasa.gov/}} (GCN) only.
Appendix~\ref{appendix:data} provides the instrument data sources and supplementary information.

Table~\ref{tab:grbs} lists the 199 Konus-\Wind short GRBs observed by the IPN. 
The first column gives the burst designation, ``GRBYYYYMMDD\_Tsssss'', where YYYYMMDD is the burst date, and sssss 
is the KW trigger time (s UT) truncated to integer seconds (note that, due to \Wind's large distance
from Earth, this trigger time can differ by up to $\sim 5.6$~s from the near-Earth spacecraft detection times). 
The second column gives the KW trigger time in the standard time format. 
The ``Name'' column specifies the GRB name as provided in the Gamma-ray Burst Coordinates Network circulars, if available.
The ``Type'' column specifies the burst type
following the classification given in~\cite{Svinkin16,Svinkin19}. 
The types are: I (merger-origin), II (collapsar-origin), I/II (the type is uncertain), 
Iee (type I with EE), and Iee/II (the type is uncertain: Iee or II). 
The ``Observed by'' column lists the missions/instruments which
observed the burst.

We found that 198 of 199 KW short GRBs were observed by at least one other IPN
spacecraft (s/c), enabling their localizations to be constrained by triangulation.
The detections are given in Table~\ref{tab:grbs} and also available on 
the IPN website\footnote{\url{http://ssl.berkeley.edu/ipn3/masterli.txt}}.
In total, 164 ($\sim 82$\%) GRBs were observed by \INT, 185 ($\sim 92$\%) by any near-Earth s/c,
and 119 ($\sim 60$\%) by distant s/c (\MO and/or \MES): 24 by two distant s/c and 95 by one distant s/c;
27 bursts were precisely localized by \Swift BAT or XRT (including GRB~150831A, which was also localized by \INT-IBIS/ISGRI).
The statistics of the events detected by each spacecraft are given in Table~\ref{tab:instruments}.

\section{LOCALIZATIONS}\label{sec:localizations}

\subsection{Triangulation Annuli}
Using the triangulation technique identical to that of P13
one or more triangulation annuli have been obtained for 198 Konus-\Wind short bursts.

\begin{figure}
\fig{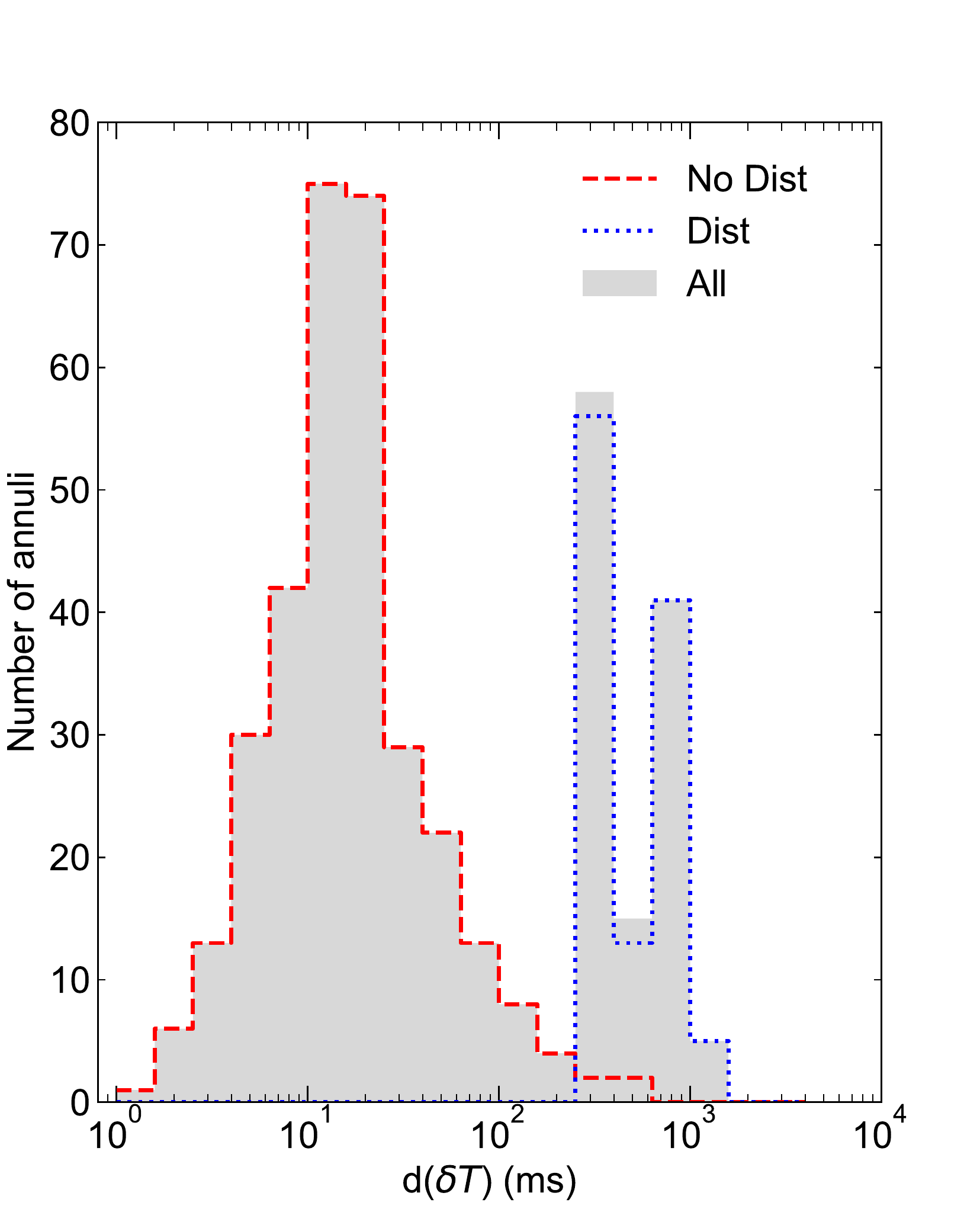}{0.48\textwidth}{a)}
\fig{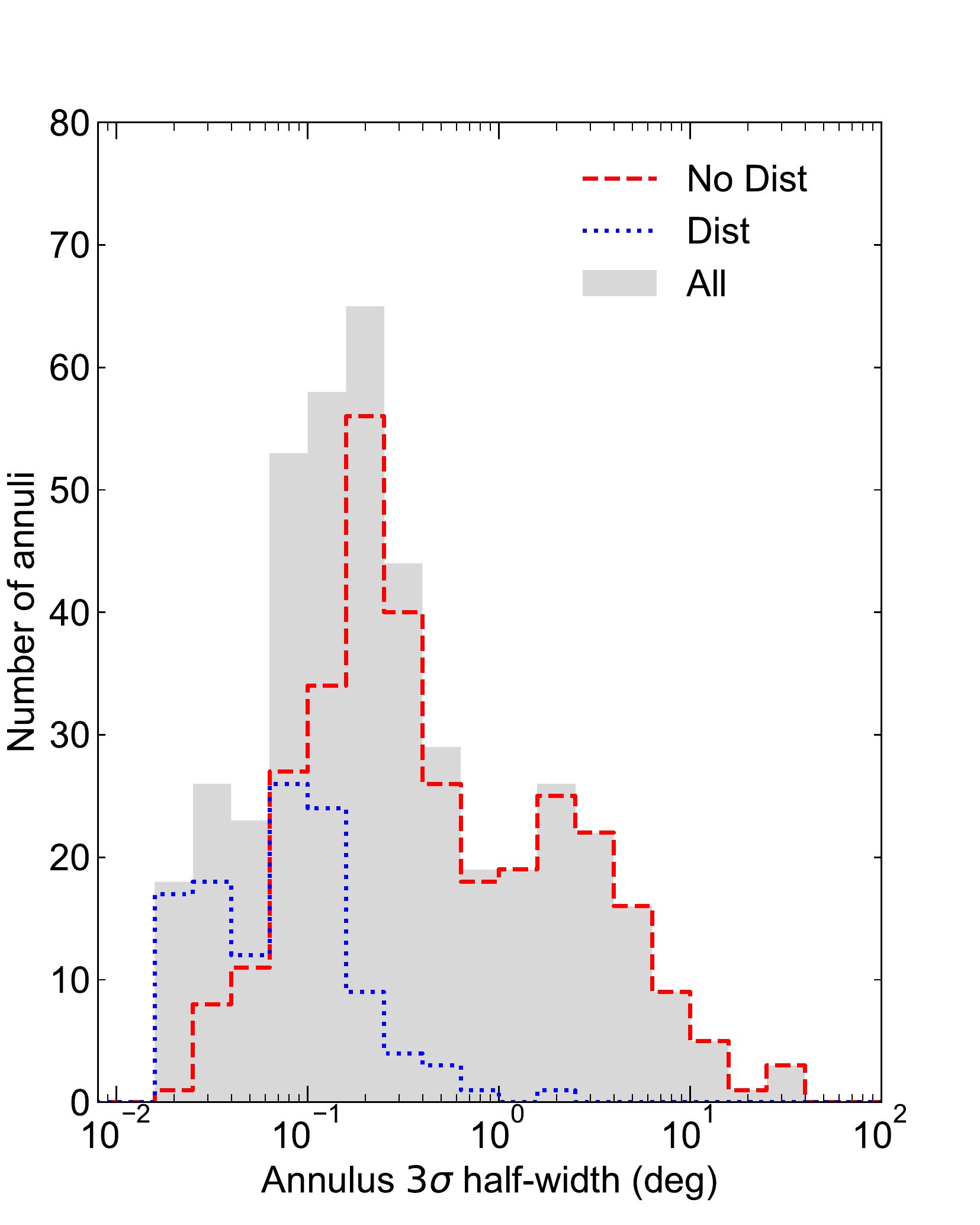}{0.48\textwidth}{b)}

\caption{Distributions of uncertainties in time delay $d(\delta T) = (d_{+}(\delta T) + |d_{-}(\delta T)|)/2$ (left) and 
$3\sigma$ half-widths (right) of the 438 triangulation annuli.
Blue dotted lines: 115~annuli involving at least one distant s/c; 
red dashed lines: 323~annuli not involving any distant s/c.
For annuli obtained using the Konus-\Wind and near-Earth (or \INT) s/c data: 
the smallest $d(\delta T)$ is 1.2~ms, the largest is 600~ms, 
the mean is 29~ms, and the geometric mean is 16~ms;
the smallest HW is 0\fdg021 (1.3\arcmin), the largest is 36\fdg7, 
the mean is 1\fdg6, and the geometric mean is 0\fdg45.
For annuli involving distant s/c:
the smallest $d(\delta T)$ is 361~ms, the largest is 1168~ms, 
the mean is 596~ms, and the geometric mean is 544~ms;
the smallest HW is 0\fdg016 (0.9\arcmin), the largest is 2\fdg1, 
the mean is 1\fdg12, and the geometric mean is 0\fdg72.
\label{fig:dtcc_errs}}
\end{figure}

From about 500 derived annuli, 435 were used in the catalog (including 115 annuli with distant s/c). 
For each burst we selected the annuli which form the smallest error region. 
Figure~\ref{fig:dtcc_errs} shows the distributions of uncertainties in time delays and $3\sigma$ half-widths of these annuli.

The detectors in the IPN vary widely in shape, composition, time resolution, and energy range.
Also, on-board timekeeping techniques and accuracies differ from mission to mission,
and spacecraft ephemeris data are given only as predicts for most of missions. 
The detailed discussion of triangulation systematic effects is given in~\citet{Hurley_2017ApJS_229_31}.
Since the accuracy of the triangulation technique depends on all these parameters, end-to-end calibrations
and sensitivity checks are a constant necessity.

\subsection{Verifying Triangulation Annuli}

Of the bursts localized by IPN, 27 were precisely localized by \Swift BAT or XRT (including GRB~150831A, which was also localized by \INT-IBIS/ISGRI).
We utilized these bursts to verify our triangulations.

For these 27 bursts, 43 KW-near-Earth s/c (including 20 KW-\INT) and 16 KW (or \Fermi)-distant s/c annuli were obtained.
The \Swift localizations were taken from either the \Swift(XRT) catalog\footnote{\url{https://www.swift.ac.uk/xrt_positions}}, 
if an X-ray afterglow was found, or the third \Swift(BAT) catalog~\citep{Lien_2016ApJ_829_7}. 
For recent GRBs, the BAT localizations were taken from GCN Circulars with refined positions.
In each case the triangulation annuli are in agreement with the precise Swift localization of the source, thereby confirming the reliability of
our triangulations. The maximum offset of $3.3\sigma$ was found for the bright GRB~130603B; 
for this burst, it corresponds to $\sim 1$~ms systematic uncertainty in time delay. 

\begin{figure}
\fig{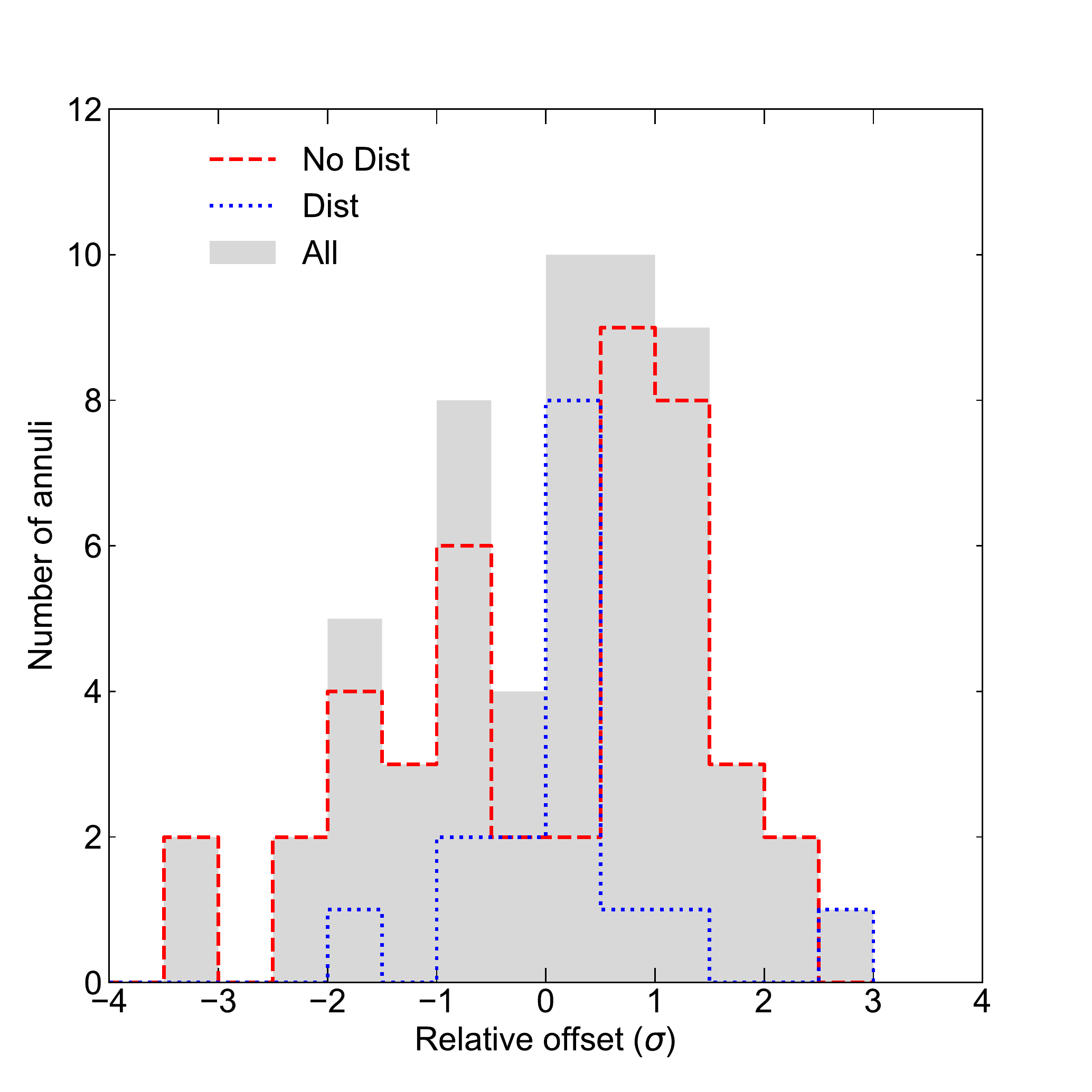}{0.5\textwidth}{}

\caption{Distributions of relative offsets (in $\sigma$) of the 27~precise GRB positions from the
center lines of the IPN annuli. The red dashed line: 43~KW-near-Earth s/c (including 20~KW-\INT) annuli; 
the blue dotted line: 16~KW (or \Fermi)-distant s/c annuli. 
\label{fig:dist_offsets}}
\end{figure}

Figure~\ref{fig:dist_offsets} shows the distributions of relative
source offsets (in sigma) from the center lines of 43 KW-near-Earth s/c (including 20 KW-\INT) and 16 KW (or \Fermi)-distant s/c annuli. 
For these subsamples, the mean offsets are 0.0 and 0.2; the standard deviations are 1.4 and 0.9, respectively.

\subsection{Additional constraints}

In addition to triangulation annuli, several other types of localization information are included in this catalog. 
They are ecliptic latitude range; autonomous burst localizations obtained by \Fermi (GBM and LAT) or GECAM; 
and Earth- or Mars-blocking (\MES was in an eccentric orbit around Mercury, so Mercury-blocking was quite rare).
This additional information helps to constrain the triangulation position, i.e., to choose one of two triangulation boxes, 
or to eliminate portions of a single annulus.
In some cases the position of the BAT coded field of view may constrain burst localization. 
In case a burst produced the significant ($\gtrsim 5\sigma$) response in the BAT summed-array rate light curve in the 15-350~keV energy band 
(see, e.g.,~\citealt{Tohuvavohu_2020ApJ_900_35} for the BAT data product description) and no BAT trigger was reported,
the burst is most probably located outside the coded field of view of the BAT.
Precise burst localizations by \Swift BAT or XRT are provided for verification of the IPN localizations.

\subsubsection{Ecliptic Latitudes}

The ecliptic latitudes of the bursts are derived by comparing the count rates of the two KW detectors
(S1 and S2) mounted on the opposite faces of the rotationally stabilized \Wind spacecraft. The axis of S2 points toward the north ecliptic pole, and the
axis of S1 points toward the south ecliptic pole.

The triggered mode data are available from a single KW detector, typically the one with a smaller GRB incidence angle ($<90^\circ$).
Intense GRBs produce count rate increases in the waiting mode time history of both KW detectors measured with 2.944~s time resolution.
The lack of a reliable \Wind mass model and the spacecraft rotation do not allow us to directly derive the \Wind incidence angle (the source ecliptic latitude) 
in a way similar to the \Fermi(GBM) and GECAM burst location techniques. 
Nevertheless, it is possible to estimate the source ecliptic latitude from the ratio of $\sim 80$--350~keV count rates in S1 and S2, 
calibrated using a sample of well localized GRBs.

The ecliptic latitude range, namely the best estimate $\beta$, and the lower and upper limits $\beta_\mathrm{min}$, $\beta_\mathrm{max}$ can be considered
to be an annulus centered at the north pole, 
with a half-angle $\theta = 90^{\circ} - \beta$ and half-width $d_{-}(\theta) =  \beta - \beta_\mathrm{max}$, $d_{+}(\theta) =  \beta - \beta_\mathrm{min}$.
The ecliptic latitude uncertainty is estimated at the 99.73\% ($3\sigma$) confidence level.

\subsubsection{Planet Blocking}\label{sec:planet_blocking}

When a spacecraft in low Earth or Martian orbit detects a burst the planet blocking may constrain burst localization, 
since the source position must be outside occulted part of the sky.
Since \Fermi(GBM) has a higher sensitivity than KW across the unocculted sky 
it is possible to use GBM non-detections to constrain GRB positions. In this case the burst source 
is inside the \Fermi Earth-occulted region. To check that GBM is switched on and collecting data we use GBM POSHIST and CTIME
data.  

The allowed part of the sky is specified in the catalog as a degenerate annulus centered at the spacecraft's nadir vector,
with a half-angle $\theta = \arcsin(R_\mathrm{planet}/R)$, where $R$ is the radius of the s/c orbit and $R_\mathrm{planet}$ 
is the solid planet radius assuming a spherical shape.
The annulus half-widths are $d_{-}(\theta) = -\theta$, $d_{+}(\theta) = 0$, in case the burst was occulted by the planet for the spacecraft 
and $d_{-}(\theta) = 0$, $d_{+}(\theta) = 180^{\circ} - \theta$ in the opposite case.

\subsubsection{Autonomous Localizations}

The autonomous localizations, derived by comparing the count rates of several detectors with a cosine-like angular response, are
affected by Earth albedo and absorption or scattering in the s/c structure, among other things; as a result, their shapes are rather complex.
To produce the final IPN localization region we use GBM RoboBA localizations~\citep{Goldstein_2020ApJ_895_40} 
in the HEALPix format publicly available for bursts since 2018; pre-2018 RoboBA localizations were provided for this work by the GBM team. 

The error circles provided in GBM GRB catalogs (e.g., \citealt{von_Kienlin_2020ApJ_893_46}), are simple approximations to these shapes. 
They are centered at the most likely arrival direction for the burst, and their radii are defined as
an average distance to the true 68\% statistical-only error contour~\citep{Connaughton_2015ApJS_216_32}.
In this catalog, in cases where GBM localizations constrain the burst location, we provide the GBM error circles for reference.
For GRB20210307\_T21404 (the only burst in the catalog detected by GECAM) we provide the GECAM localization. 

\Fermi(LAT) localizations provided in the catalog are circles centered at the LAT best reconstructed position 
with 90\% containment radius (statistical only) were taken from either the \Fermi(LAT) catalog~\citep{Ajello_2019ApJ_878_52}
or, for recent GRBs, from GCN Circulars.

\subsection{Localization regions}

\begin{figure}

\includegraphics[width=0.9\textwidth]{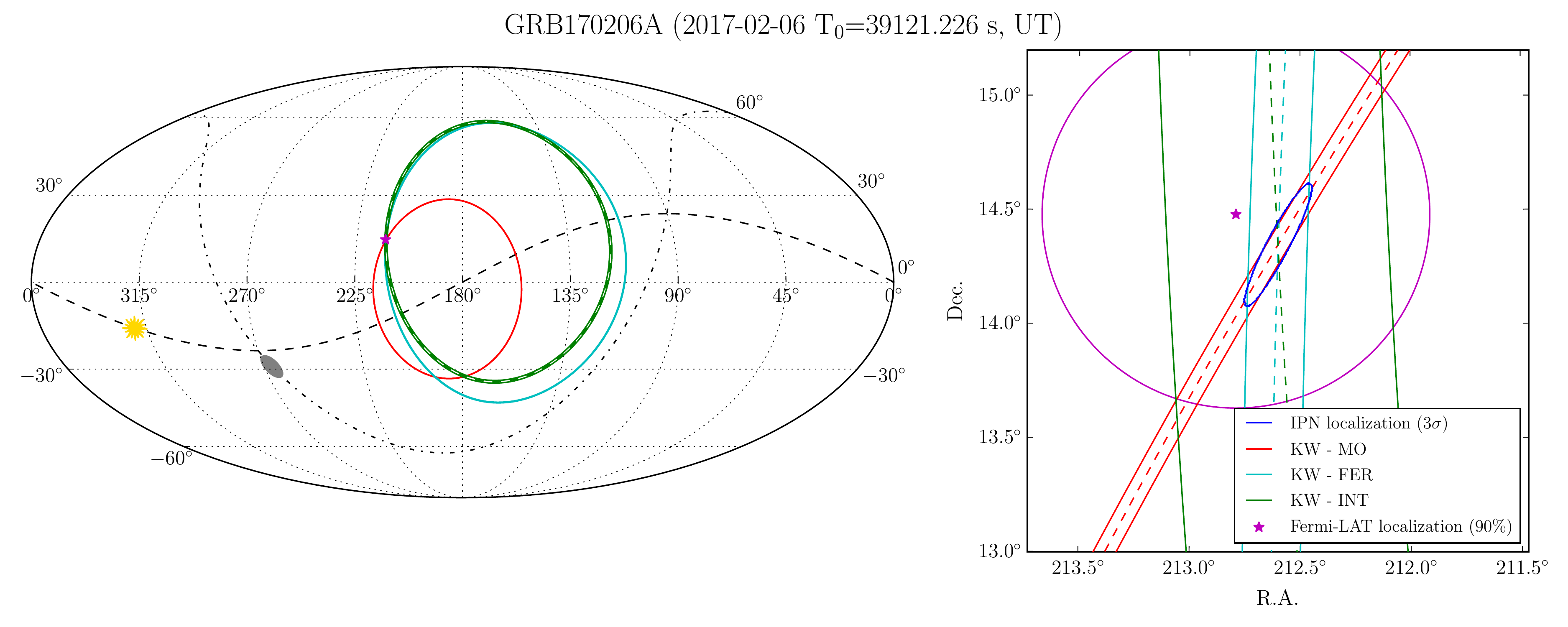}
\includegraphics[width=0.9\textwidth]{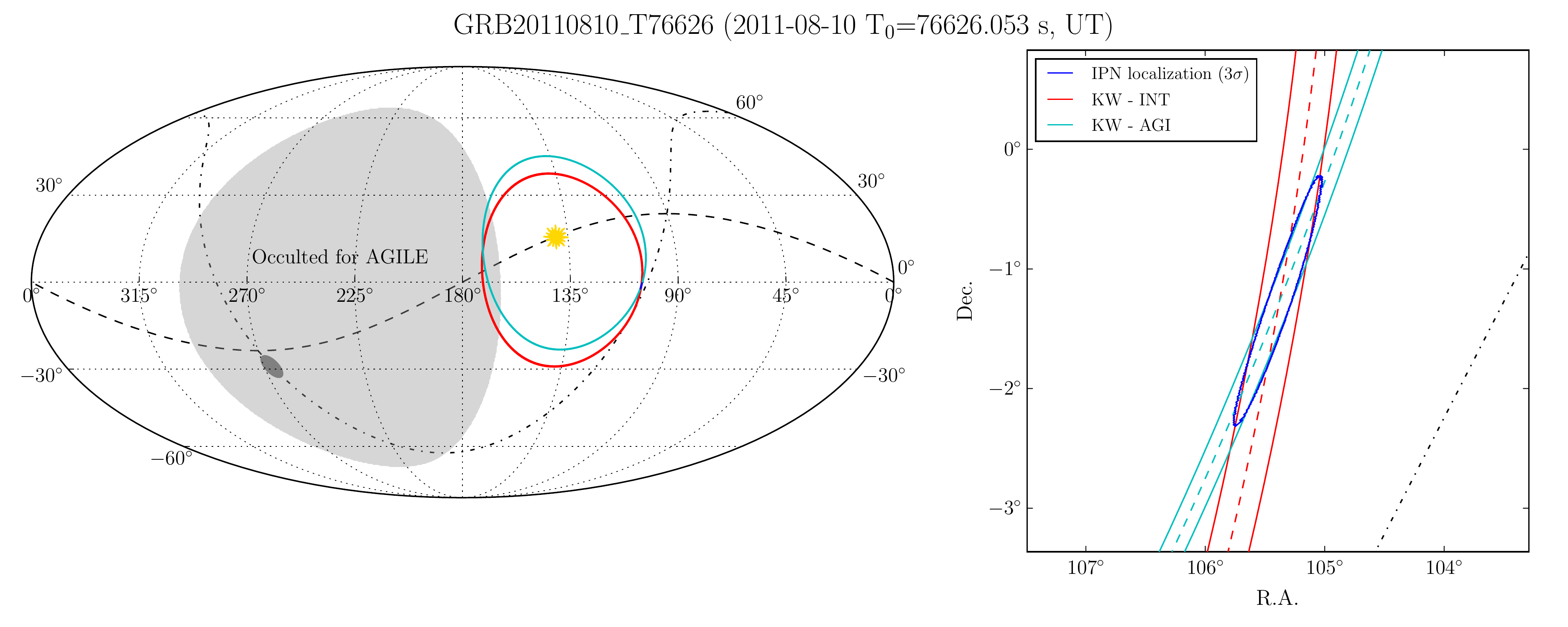}
\includegraphics[width=0.9\textwidth]{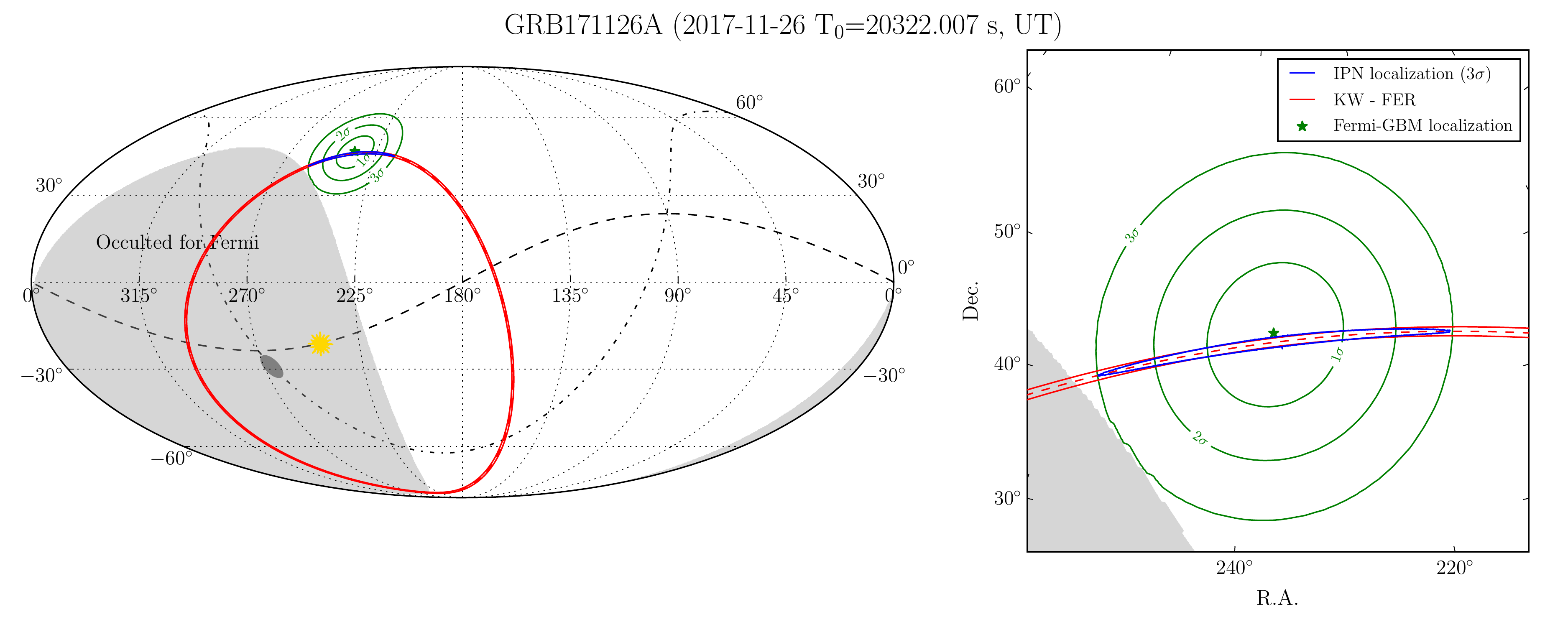}
\caption{IPN localizations. 
Three panels present typical cases of IPN localizations:
GRB detected by four well separated s/c (top panel); 
GRB detected by three s/c with an additional constraint used to select the final localization (middle panel);
GRB detected by two s/c, the localization is formed by a single annulus and GBM localization (bottom panel).
The left plot in each panel shows the whole sky with triangulation annuli and other constraints along with
the ecliptic plane (dashed line) and Sun position at the GRB time;
the Galactic plane (dash-dotted line) and the Galactic center.
The right plots are magnification insets showing the $3\sigma$ localization confidence region (blue solid line).
The GRB name and the KW trigger time are given in the figure title.
\label{fig:loc_figure}
}
\end{figure}

The final localizations were produced using the set of IPN annuli and additional constraints.
For those bursts which were detected by three or more well separated s/c, 
a small localization region (down to tens of arcmin$^2$) can be derived (Figure~\ref{fig:loc_figure}, top panel).
For such bursts, we have typically provided three annuli in the catalog. 
We used the two of them which provided the most compact localization region, to construct the localization. 
As a typical result two regions are produced and, with the help of the third annulus, we selected the final region.

For bursts not observed by any distant s/c, but observed by KW, \INT(SPI-ACS), 
and one or more near-Earth s/c, the localization region is formed by the intersections of the KW-near-Earth s/c annulus and 
an \INT-near-Earth s/c annulus, or by a KW-near-Earth s/c annulus and a KW-\INT annulus intersecting at grazing incidence.
In this case the final region was selected taking into account additional constraints (Figure~\ref{fig:loc_figure}, middle panel). 
Where a GBM localization is used we exclude a region if it lies outside the GBM $3\sigma$ contour (calculated using the RoboBA localization).
We also used \Fermi(LAT) and precise \Swift localizations (where available) for region selection.

For those bursts which were detected only by KW and one other s/c, or by KW and one or more near-Earth s/c, the resulting
localization is formed by a triangulation annulus (the narrowest in the case of several KW-near-Earth s/c annuli) and additional
constraints. These localizations consist of the entire annulus (in the case where it is entirely inside the allowed ecliptic latitude
band and there are no other constraints) or one or two annulus segments, formed by the intersection of the annulus with the
ecliptic latitude band, and/or by exclusion of the occulted part of the annulus, or by combination with the GBM localization 
(Figure~\ref{fig:loc_figure}, bottom panel).

\subsubsection{Probability sky maps}
The localization maps were produced in the multi-resolution map HEALPix format\footnote{\url{https://www.ivoa.net/documents/MOC}}
using the \texttt{mhealpy} Python package\footnote{\url{https://mhealpy.readthedocs.io}} in the following way.
The probability density for the annulus was specified by a Gaussian distribution centered at the symmetrized annulus center line and 
having $3\sigma$ half-width equal to the annulus half-width. 
Localization annuli are typically asymmetrical with respect to the annulus center line; 
they are symmetrized by defining an average annulus half-width about a displaced center line.
The planet blocking regions were specified by a degenerate annulus on the sky with a uniform probability inside it and zero probability outside, 
see Section~\ref{sec:planet_blocking}.

In case a burst was detected by BAT outside the coded field of view, the BAT coded field of view was represented as 
a region with partial coding fraction $>10$\% with zero probability inside it.
The KW ecliptic latitude band was represented as an annulus with a uniform probability inside it and zero probability outside, 
centered at the north ecliptic pole with radius and half-width corresponding to the incident angle constraint.
The final probability density was calculated as a product of the selected annuli and constraints.

We note that, in this work, the probability densities for each annuli are assumed to be
independent, despite that they may involve overlapping data (e.g. in case of intersection of KW-FER and FER-INT annuli). 
Such simplification can be avoided if future with more sophisticated methods (e.g.~\citealt{Burgess_2021AandA}).

The localization contours were calculated using the \texttt{ligo.skymap} package\footnote{\url{https://lscsoft.docs.ligo.org/ligo.skymap}}, 
modified for the multi-resolution map case. Examples of the localization maps are given in Figure~\ref{fig:loc_figure}.
The probability sky maps are stored in files following the 
Flexible Image Transport System (FITS) standard (see Appendix~\ref{appendix:loc} for file description).

\section{LOCALIZATIONS: RESULTS}\label{sec:loc_res}
Table~\ref{tab:loc} summarizes localization information for 199 Konus short bursts. 
The first column gives the burst designation (see Table~\ref{tab:grbs}).
The second column gives the number of localization constraints (the number of rows with localization information for the burst). 
The six subsequent columns give localizations expressed as a set of annuli: 
the third column gives the source of the location: either sc1-sc2 (triangulation annulus derived using sc1 and sc2), 
or ``Ecl.Lat'' (range of ecliptic latitudes), or ``Pos.Instr'' (`Instr' is one of the following: 
`SWI' for \Swift BAT or XRT, `GBM' or `LAT' for \Fermi, and `GEC' for GECAM localizations), or 
``Occ.sc'' (planet blocking for sc); Columns 4-8 list the right ascension and declination of the annulus center,
the annulus radius $\theta$, and the $3\sigma$ uncertainties in the radius $d_{-}(\theta)$, $d_{+}(\theta)$.
Planet blocking, ecliptic latitude range, and autonomous localizations are given only if they constrain the location.

\begin{figure}
\includegraphics[width=0.45\textwidth]{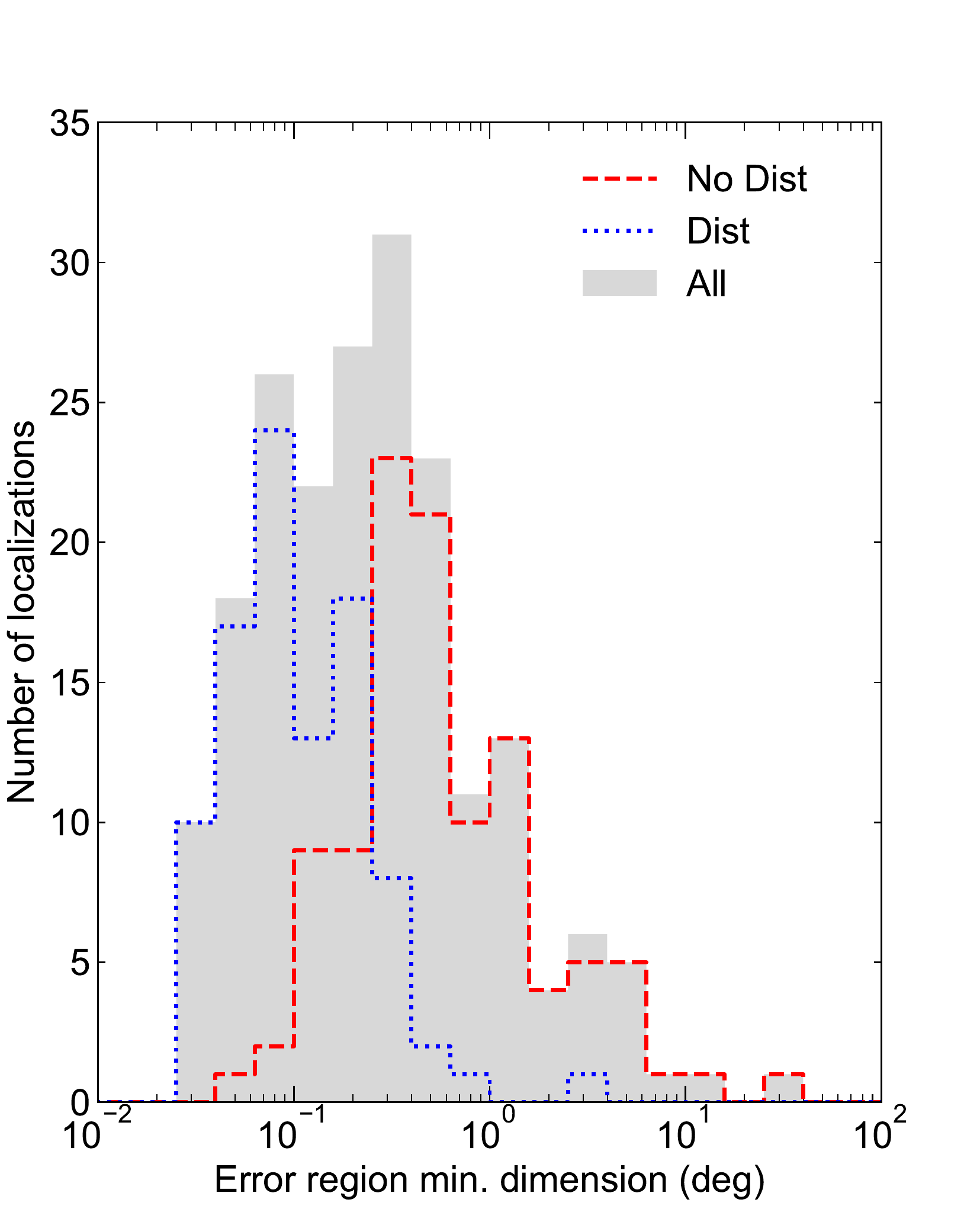}
\includegraphics[width=0.45\textwidth]{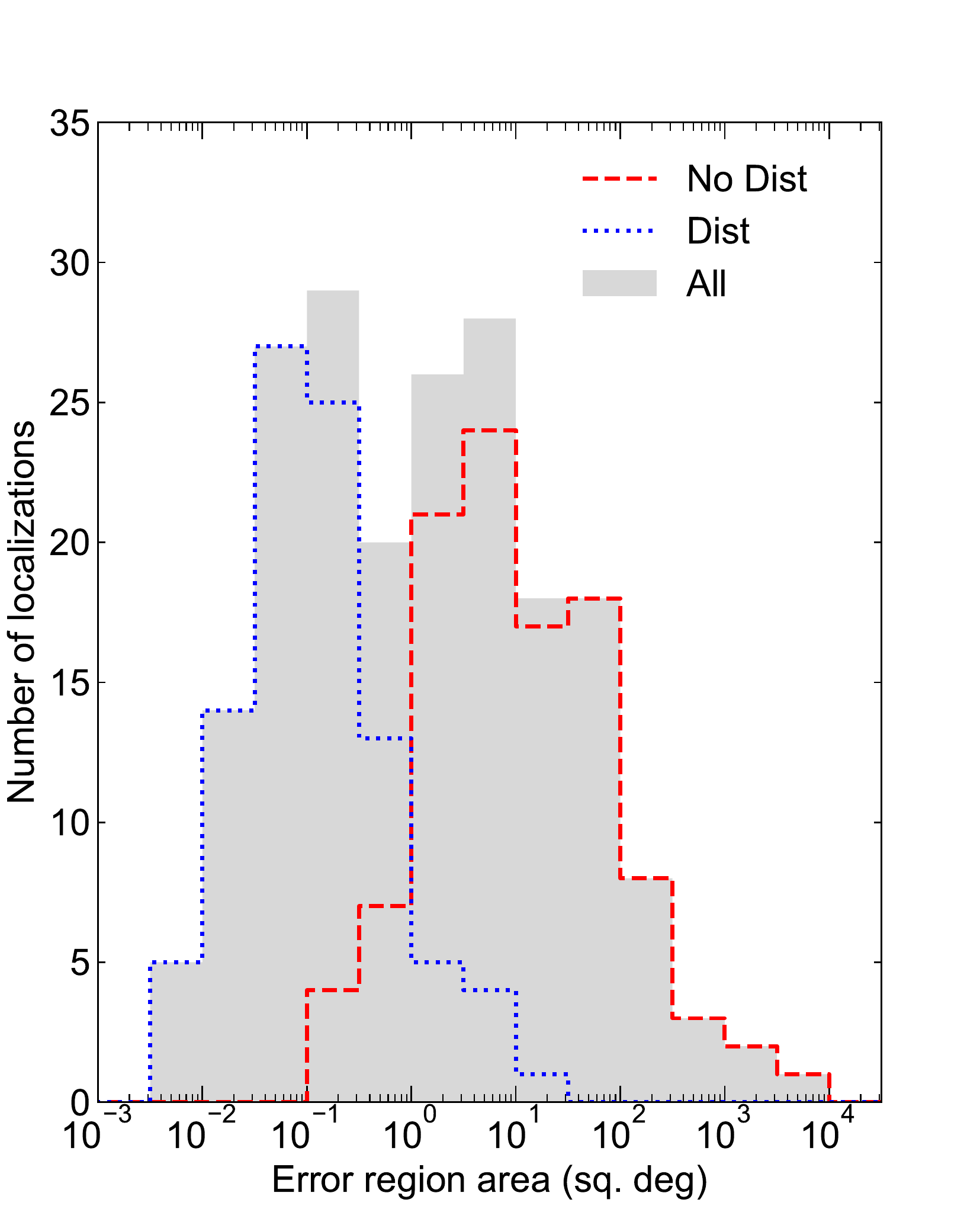}
\caption{Distributions of minimum sizes (left) and areas (right) of the IPN localizations. 
Blue dotted lines: 94~Konus short bursts observed by at least one distant s/c; 
red dashed lines: 105~bursts not observed by any distant s/c.
The minimum region dimensions range  
from 0.033~deg (2.0~arcmin) to 3.32~deg with a mean of 0.17~deg, and a geometric mean of 0.11~deg (for bursts observed by distant s/c) and 
from 0.043~deg (2.6~arcmin) to 36.5~deg with a mean of 1.5~deg, and a geometric mean of 0.6~deg (for bursts without distant s/c detections).
The areas range
from 0.005~deg$^2$ (18~arcmin$^2$) to 14.4~deg$^2$ with a mean of 0.70~deg$^2$, and a geometric mean of 0.12~deg$^2$ (for bursts observed by distant s/c) and 
from 0.156~deg$^2$ to 3163~deg$^2$ with a mean of 103~deg$^2$, and a geometric mean of 11.0~deg$^2$ (for bursts without distant s/c detections).
\label{fig:region_sizes}}
\end{figure}

Table~\ref{tab:regions} gives the description of the final IPN error regions (including the 27 imaged bursts). 
The nine columns contain the following information: 
(1) the burst designation (see Table~\ref{tab:grbs}); 
(2) the number of error regions for the burst, $N_r$: 1 or 2; 
(3) and (4) the right ascension and declination of the most probable burst location for each region;
(5) the maximum dimension of the region (that is, the maximum angular distance between two points at the 99.73\% probability region boundary);
(6) the minimum dimension of the region (that is, the full width of the narrowest annulus forming the region);
(7) the area (for two regions the area of each region) enclosing 99.73\% probability. 
Distributions of the region dimensions and areas are shown in Figure~\ref{fig:region_sizes}.

\section{CONCLUSIONS}\label{sec:conclusions}
This paper continues a series of catalogs of GRB localizations obtained by arrival-time analysis, 
or ``triangulation'' between the spacecraft in the third IPN, as summarized in Table~\ref{tab:IPN_catalogs}.
We have presented the most comprehensive IPN localization data on 199 Konus-\Wind short bursts detected between 2011 January~1 and 2021 August~31. 
With one exception, IPN localizations were obtained for these events (for GRB20171108\_T51656 observed by KW only, 
the source position is constrained to the \Fermi Earth-occulted region combined with the KW ecliptic latitude range). 
We verified the triangulations using 27 bursts localized by instruments with imaging capability. 
In each case the derived IPN annuli are in agreement with the precise GRB position, 
thereby confirming the reliability of our results.

Currently the nine-spacecraft IPN detects about 325 bursts per year~\citep{Hurley_2017ApJS_229_31}, 
about 18 of which are rather bright, short-duration, hard spectrum GRBs (see P13 and this work).
The IPN localizations can be used for a wide variety of purposes, including, but not limited to,
searches for gravitational wave, kilonovae, and neutrino signals from merging compact objects, 
very high energy photons from the burst sources, and giant magnetar flares in nearby galaxies.
As Konus-\Wind continues to operate, we anticipate more localizations, in particular in conjunction 
with upcoming LIGO/Virgo operations.

\begin{acknowledgments}

DS, AR, AL, DF, and MU acknowledge support from  RSF grant 21-12-00250.
The HEND experiment was supported by the Russian State Corporation Roscosmos and 
implemented as part of Gamma-Ray Spectrometer suite on NASA \MO. 
HEND data processing was funded by Ministry of Science and Higher Education of the Russian Federation, grant AAAA-A18-118012290370-6.
KH is grateful for support under the Fermi Guest Investigator program, grant~80NSSC20K0585.
We thank Valentin Pal'shin for his considerable contribution to the Konus-\Wind and IPN data analysis tools.
Some of the results in this paper have been derived using the healpy and HEALPix packages.
\end{acknowledgments}

\vspace{5mm}
\facilities{
\Wind(Konus), \Fermi(GBM and LAT), \Swift(BAT and XRT), \Suzaku(WAM), \AGILE(MCAL),
\emph{ISS}(CALET-CGBM), \Insight-HXMT(HE), GECAM-B, \INT(SPI-ACS), \MES(GRNS), \MO(HEND)
}

\software{
Astropy \citep{Astropy_2013AandA, Astropy_2018AJ};
\Fermi GBM Data Tools~\citep{GbmDataTools};
Utilities for \Swift(BAT) instrument (\url{https://github.com/lanl/swiftbat\_python});
Astroquery~\citep{Ginsburg_2019AJ_157_98};
healpy (\citet{Gorsky_2005ApJ_622_759, Zonca_2019JOSS_4_1298}, \url{https://healpix.sourceforge.io}); 
mhealpy (\url{https://mhealpy.readthedocs.io}); 
ligo.skymap (\url{https://lscsoft.docs.ligo.org/ligo.skymap}).
}

\appendix

\section{DATA SOURCES}\label{appendix:data}
\subsection{IPN instrument data}
We use the following sources for the instrument data: 
\Swift-BAT (\url{https://heasarc.gsfc.nasa.gov/FTP/swift/data/obs/}), 
for recent GRBs we used BAT time-tagged event data that exist due to the Gamma-ray Urgent Archiver for
Novel Opportunities (GUANO; \citealt{Tohuvavohu_2020ApJ_900_35}).
\Fermi-GBM  (\url{https://heasarc.gsfc.nasa.gov/FTP/fermi/data/gbm/triggers/});
\Suzaku-WAM  (\url{http://www.astro.isas.jaxa.jp/suzaku/HXD-WAM/WAM-GRB/grb/trig/grb_table.html});
\INT-SPI-ACS (\url{http://isdc.unige.ch/~savchenk/spiacs-online/});
\MES-GRNS (\url{https://pds-geosciences.wustl.edu/missions/messenger/grns.htm});
\AGILE-MCAL, \Insight-HXMT, and \MO-HEND data are available on request from the instrument teams.
The \AGILE-MCAL short GRB data are part of the the Second AGILE MCAL GRB Catalog (Ursi~et~al., submitted to ApJ, 2021).

\subsection{Ephemeris and clock accuracy data}
Near-Earth spacecraft ephemerides were derived from two-line elements (TLE) available at \url{https://www.space-track.org} using SGP8 model. 
For \Wind we use the predicted ephemeris; for the \MO and \MES ephemerides, and the \MO nadir vectors, 
we used the JPL Horizons on-line ephemeris system (\url{https://ssd.jpl.nasa.gov/horizons.cgi})
using the \texttt{astroquery} Python package~\citep{Ginsburg_2019AJ_157_98}.
\INT ephemeris data are available via the SPI-ACS data web-interface.
The \Wind predicted ephemeris data and their description
are available at \url{https://spdf.gsfc.nasa.gov/pub/data/wind/orbit/pre\_or/} and
\url{https://cdaweb.gsfc.nasa.gov/misc/NotesW.html\#WI\_OR\_PRE}.

The declared on-board clock accuracy of the spacecraft are: down to 1~$\mu$s for \Fermi; $\sim 200$~$\mu$s for
\Swift; $\sim 1$~ms for \Wind; and $\sim 100$~$\mu$s for \INT; for \MO an overall $3\sigma$  systematic uncertainty,
which includes timing and other effects derived from IPN observations of precisely localized GRBs, is
better than 360~ms, the corresponding \MES uncertainty is~800~ms. 
The \Swift timing corrections were calculated using Swift-BAT utilities.
The Wind clock drift information is provided at \url{ftps://pwgdata.sci.gsfc.nasa.gov/pub/wind_clock/}.

\section{LOCALIZATION FILES}\label{appendix:loc}
The localizations are stored in a series of FITS data files available on-line from the Ioffe web-site\footnote{\url{http://www.ioffe.ru/LEA/ShortGRBs_IPN/}}
in the following formats:
multi-resolution HEALPix maps (\texttt{*\_IPN\_map\_hpx\_moc.fits.gz}),
regular resolution HEALPix maps (\texttt{*\_IPN\_map\_hpx.fits.gz}).
In addition to HEALPix maps the web-site stores 
the localization 0.9973 integrated probability contours in ASCII format
which contain the coordinates of the center of the pixel with the maximum probability (two pixels are given in the case of two localization regions), 
followed by the region contour coordinates;
and plots produced using the \texttt{ligo.skymap} package.

The regular resolution maps are standard HEALPix FITS files with RING numeration, coordinate system~-- celestial.
The single extension table of the file contains the pixel probability.
The comment field contains information about the instruments involved in localization, a list of additional constraints,
and the parameters of the IPN annuli used. 
The extension table of the multi-resolution map file contains the pixel probability and the pixel index in the  
UNIQ indexing scheme\footnote{\url{https://emfollow.docs.ligo.org/userguide/tutorial/multiorder_skymaps.html}}~\citep{Singer_2016PhRvD_93b4013}. 

\newpage

\clearpage

\input{table_1_IPN.tex}
\input{table_2_GRBs.tex}
\input{table_3_annuli.tex}
\input{table_4_regions.tex}
\input{table_5_prev_IPN_cat.tex}

\clearpage

\bibliographystyle{aasjournal}
\bibliography{bibliography}

\end{document}

%% file: table_1_IPN.tex
\begin{deluxetable*}{lccccccc}
    \tabletypesize{\scriptsize}
    \tablewidth{0pt}
    \tablecaption{IPN composition in 2011-2021 \label{tab:instruments}}
    \tablehead{   
        \colhead{Mission (Instrument)\tablenotemark{a}} \vspace{-0.2cm}%
        & \colhead{Designation}%
        & \colhead{Operation}%
        & \colhead{Earth distance\tablenotemark{b}}%
        & \colhead{Energy band\tablenotemark{c}}%
        & \colhead{Time resolution\tablenotemark{d}}%
        & \colhead{N$_\mathrm{GRBs}$\tablenotemark{e}}\\
        \colhead{}%
        & \colhead{}%
        & \colhead{period}%
        & \colhead{(lt.~sec)}%
        & \colhead{(keV)}%
        & \colhead{(ms)}%
        & \colhead{}%
    }
    \startdata
    \emph{Wind}~(Konus)       & KW    & since 1994  & up to $\sim 6$    & 80--1500       & 2--256 (T)      & 199  \\
    \emph{INTEGRAL}~(SPI-ACS) & INT   & since 2002  & up to $\sim 0.5$  & 75--8000       & 50  (R)         & 164  \\
    \emph{Swift}~(BAT)        & SWI   & since 2004  & LEO               & 25--350        & 64 (R), TTE (T) & 128  \\
    \emph{Fermi}~(GBM)        & FER   & since 2008  & LEO               & 80--1000       & TTE (T)         & 113  \\
    \emph{AGILE}~(MCAL)       & AGI   & since 2007  & LEO               & $\gtrsim 400$  & TTE (T)         & 52   \\
    \emph{Suzaku}~(WAM)       & SUZ   & 2005--2015  & LEO               & 110--5000      & 1/64 s (T)      & 37   \\
    \emph{Insight}-HXMT~(HE)  & INS   & since 2017  & LEO               & 200--3000      & TTE (T)         & \textbf{35}   \\
    \emph{RHESSI}~(GeD)       & RHE   & 2002--2018  & LEO               & $\gtrsim 100$  & TTE             & 27   \\
    \emph{ISS}~(CALET-CGBM)   & CAL   & since 2015  & LEO               & $\gtrsim 40$   & TTE (T)         & 15   \\
     GECAM-B~(GRD)            & GEC   & since 2020  & LEO               & 10--5000       & TTE (T)         & 1    \\
    \emph{Mars-Odyssey}~(HEND)& MO    & since 2001  & up to $\sim 1250$ & 50--3000       & 250 (R)         & 72   \\
    \emph{MESSENGER}~(GRNS)   & MES   & 2004--2015  & up to $\sim 700$  & 40--200        & 1000 (T)        & 47   \\
    \enddata  
    \tablenotetext{a}{%
        Instruments providing burst localizations, but not used for the triangulation:
        \emph{Fermi} Large Area Telescope (LAT); %
        \emph{Swift} X-ray telescope (XRT); %
    }
    \tablenotetext{b}{Light travel time from the s/c to the Earth center; LEO: low Earth orbit.}
    \tablenotetext{c}{Energy range used for triangulations}
    \tablenotetext{d}{
        TTE stands for time-tagged event data.
        In parentheses the detection mode is given: T~--- trigger, R~--- waiting mode rate increase.}
    \tablenotetext{e}{
        Number of Konus short bursts observed by each mission (for KW~--- the total number of bursts is given).
    }
   
\end{deluxetable*}

%% file: table_2_GRBs.tex
\begin{deluxetable}{ccccrr}
	\tablewidth{0pt}
	%
	%
	\tablecaption{IPN/Konus short gamma-ray bursts  \label{tab:grbs}}
	%
	%
	\tablehead{
		\colhead{Designation}%
		& \colhead{Konus-\emph{Wind}}%
        & \colhead{Name\tablenotemark{a}}%
		& \colhead{Type}%
		& \colhead{Observed by\tablenotemark{b}}\\
		\colhead{}%
		& \colhead{trigger time (UT)}%
        & \colhead{}%
		& \colhead{}%
		& \colhead{}
    }
	\startdata
    GRB20110212\_T47551  & 13:12:31.101 & \nodata    & I    &  INT(R),SWI(R),SUZ(T),AGI(T),FER(T) \\
    GRB20110221\_T18490  & 05:08:10.017 & \nodata    & I    &   MO(R),RHE(R),INT(R),SUZ(T),AGI(T) \\
    GRB20110323\_T57460  & 15:57:40.228 & \nodata    & I    &                       INT(R),SUZ(T) \\
    GRB20110401\_T79461  & 22:04:21.937 & GRB~110401A & I/II &         INT(R),SWI(R),AGI(T),FER(T) \\
    GRB20110510\_T80844  & 22:27:24.326 & \nodata    & I/II &                 MES(T),MO(R),SWI(R) \\
	\enddata
    \tablenotetext{a}{As provided in the GCN circulars, if available.}
	\tablenotetext{b}{%
		AGI: \emph{AGILE} (MCAL); %
        CAL: \emph{International Space Station} CALET Gamma-ray Burst Monitor;
        GEC: GECAM-B~(GRD);
		FER: \emph{Fermi}~(GBM); 
        INS: \emph{Insight}-HXMT (HE);
		INT: \emph{INTEGRAL}~(SPI-ACS); %
        KON: \Wind~(Konus);
		LAT: \emph{Fermi}~(LAT); %
		MES: \emph{MESSENGER}~(GRNS);
		MO:  \emph{Mars-Odyssey}~(HEND); %
		RHE: \emph{RHESSI}~(GeD); %
		SUZ: \emph{Suzaku}~(WAM); %
		SWI: \emph{Swift}~(BAT); %
		In parentheses the detection mode is given: T -- trigger, R -- rate increase.
    }
	\tablecomments{This table will be published in its entirety in the
		electronic edition of the Astrophysical Journal Supplement Series. A
		portion is shown here for guidance regarding its form and content.}
\end{deluxetable}

%% file: table_3_annuli.tex
\begin{deluxetable}{lccccccc}
    \tablewidth{0pt}
	\tablecaption{IPN Localization Data  \label{tab:loc}}
	%
	%
	\tablehead{
		\colhead{Designation}%
		& \colhead{$N$}%
		& \colhead{Location}%
		& \colhead{R.A.}%
		& \colhead{Decl.}%
		& \colhead{$\theta$}%
		& \colhead{$d_{-}(\theta)$}%
		& \colhead{$d_{+}(\theta)$} \\
		  \colhead{}%
		& \colhead{}%
		& \colhead{Source}%
		& \colhead{(deg)}%
		& \colhead{(deg)}%
		& \colhead{(deg)}%
		& \colhead{(deg)}%
		& \colhead{(deg)}%
    }
	\startdata
    GRB20110212\_T47551  & 3 &  KW-FER  & 326.1161 & -19.9156 & 54.4586 & -0.0677 & +0.0676 \\
                         &   &  KW-INT  & 324.7813 & -25.3330 & 49.3296 & -0.2804 & +0.1747 \\
                         &   &  Pos.GBM & 311.330  & -74.500  &  4.33   &         &         \\
    GRB20110221\_T18490  & 3 &  KW-MO   & 330.8868 & -13.0377 & 16.0404 & -0.0657 & +0.0692 \\
                         &   &  KW-SUZ  & 343.4756 & -13.1519 & 27.6975 & -0.2354 & +0.4076 \\
                         &   &  KW-INT  & 342.3842 & -20.1351 & 29.0384 & -0.1671 & +0.3316 \\                                     
    GRB20110323\_T57460  & 2 &  KW-SUZ  &  24.7970 &   9.8654 & 20.4803 & -0.4630 & +0.1954 \\
                         &   &  KW-INT  &  24.5553 &   5.7015 & 17.2516 & -0.4805 & +0.9990 \\ 
    GRB20110401\_T79461  & 3 &  KW-INT  & 211.5140 & -12.0493 & 63.8888 & -0.2219 & +0.1969 \\
                         &   &  FER-INT &  91.5378 &  68.7932 & 85.6387 & -2.1886 & +3.1509 \\
                         &   &  Occ.SWI &  115.940 &   20.532 &  66.440 & -0.0    & +113.560  \\
    GRB20110510\_T80844  & 3 &  KW-SWI  & 228.2411 & -23.4898 & 84.5588 & -0.2262 & +0.1233 \\
                         &   &  KW-MO   &  27.6375 &  10.6619 & 73.0224 & -0.0246 & +0.0246 \\
                         &   &  KW-MES  &  22.7559 &   6.0257 & 66.3059 & -0.1203 & +0.1179 \\
    \enddata

\end{deluxetable}

%% file: table_4_regions.tex
\begin{deluxetable}{lcccrcc}
	\tablewidth{0pt}
	\tablecaption{IPN Error Regions\label{tab:regions}}
	%
	%
	\tablehead{
		\colhead{Designation}%
		& \colhead{$N_r$}
		& \colhead{R.A.}
		& \colhead{Decl.}%
		& \colhead{Max. Dim.}%
        & \colhead{Min. Dim.}%
		& \colhead{Area}\\
		\colhead{}%
        & \colhead{}%
		& \colhead{(deg)}%
		& \colhead{(deg)}%
        & \colhead{(deg)}%
		& \colhead{(deg)}%
		& \colhead{(deg$^2$)}
    }
	\startdata
    GRB20110212\_T47551    &   2  &  271.471   &   -57.147   &    12.745    &    0.135   &     1.534  \\
                           &      &  348.236   &   -72.931   &    12.745    &    0.135   &     1.534  \\
    GRB20110221\_T18490    &   1  &  316.305   &    -5.895   &     3.406    &    0.135   &     0.413  \\
    GRB20110323\_T57460    &   2  &   10.195   &    -4.350   &    12.197    &    0.658   &     6.496  \\
                           &      &   37.705   &    -5.885   &    12.197    &    0.658   &     6.494  \\
    GRB20110401\_T79461    &   1  &  264.968   &   +24.902   &     6.948    &    0.419   &     2.955  \\
    GRB20110510\_T80844    &   1  &  334.020   &   -43.873   &     3.216    &    0.050   &     0.131  \\
    \enddata

\end{deluxetable}

%% file: table_5_prev_IPN_cat.tex
\begin{deluxetable}{ccc}
\tabletypesize{\scriptsize}
\tablecaption{IPN catalogs of gamma-ray bursts to date\label{tab:IPN_catalogs}}
\tablewidth{0pt} 
\tablehead{ \colhead{Years covered} & \colhead{Number of GRBs} & \colhead{Description} 
} 
\startdata
1990--1992 & 16            & \textit{Ulysses, Pioneer Venus Orbiter,} WATCH, SIGMA, PHEBUS GRBs\tablenotemark{a} \\
1990--1994 & 56            & \textit{Granat-}WATCH supplement\tablenotemark{b} \\
1991--1992 & 37            & \textit{Pioneer Venus Orbiter, Compton Gamma-Ray Observatory, Ulysses} GRBs\tablenotemark{c} \\
1991--1994 & 218           & BATSE 3B supplement\tablenotemark{d} \\
1991--2000 & 211           & BATSE untriggered burst supplement\tablenotemark{e} \\
1992--1993 & 9             & \textit{Mars Observer} GRBs\tablenotemark{f} \\
1994--1996 & 147           & BATSE 4Br supplement\tablenotemark{g} \\
1994--2010 & 279           & First Konus short bursts\tablenotemark{h} \\
1996--2000 & 343           & BATSE 5B supplement\tablenotemark{i} \\
1996--2002 & 475           & \textit{BeppoSAX} supplement\tablenotemark{j} \\
2000--2006 & 226           & HETE-2 supplement\tablenotemark{k} \\
2008--2010 & 146           & First GBM supplement\tablenotemark{l} \\
2010--2012 & 165           & Second GBM supplement\tablenotemark{m} \\
2011--2021 & 199           & Second Konus short bursts\tablenotemark{n} \\
\enddata
\tablenotetext{a}{\citet{Hurley_2000ApJ_533_884};}
\tablenotetext{b}{\citet{Hurley_2000ApJS_128_549};}
\tablenotetext{c}{\citet{Laros_1998ApJS_118_391};}
\tablenotetext{d}{\citet{Hurley_1999ApJS_120_399};}
\tablenotetext{e}{\citet{Hurley_2005ApJS_156_217};}
\tablenotetext{f}{\citet{Laros_1997ApJS_110_157};}
\tablenotetext{g}{\citet{Hurley_1999ApJS_122_497};}
\tablenotetext{h}{\citet{Palshin13};}
\tablenotetext{i}{\citet{Hurley_2011ApJS_196_1};}
\tablenotetext{j}{\citet{Hurley_2010ApJS_191_179};}
\tablenotetext{k}{\citet{Hurley_2011ApJS_197_34};}
\tablenotetext{l}{\citet{Hurley_2013ApJS_207_39};}
\tablenotetext{m}{\citet{Hurley_2017ApJS_229_31};}
\tablenotetext{n}{Present catalog.}
\end{deluxetable}